\documentclass[useAMS,usenatbib,fleqn]{mn2e}
\usepackage{aas_macros}
\usepackage{graphicx}
\graphicspath{{/murphy/seb_stuff/SFXT/COMB_OBS_ANALYSIS/IGRJ17544/}}
\usepackage{fixltx2e}
\usepackage{amsmath}
\title[Accretion in IGR J17544$-$2619]{New insights on accretion in Supergiant Fast X-ray Transients from XMM-Newton and INTEGRAL observations of IGR J17544$-$2619}
\author[S. P. Drave et al.]
	{S. P. Drave$^1\thanks{sd805@soton.ac.uk}$, A. J. Bird$^1$, L. Sidoli$^2$, V. Sguera$^{3}$, A. Bazzano$^4$, A. B. Hill$^{5,1}$
        \newauthor
         and M. E. Goossens$^1$  \\
        $^1$School of Physics and Astronomy, Faculty of Physical Sciences and Engineering, University of Southampton, University Road, \\ 
        Southampton, SO17 1BJ, UK \\
        $^2$INAF-IASF, Istituto di Astrofisica Spaziale e Fisica Cosmica, Via E. Bassini 15, I-20133 Milano, Italy \\
        $^3$INAF-IASF, Istituto di Astrofisica Spaziale e Fisica Cosmica, Via Gobetti 101, Bologna, Italy \\
        $^4$IAPS-INAF, Istituto di Astrofisica Spaziale e Fisica Cosmica, Via del Fosso del Cavaliere 100, 00133 Roma, Italy \\
        $^5$W. W. Hansen Experimental Physics Laboratory, Kavil Institute for Particle Astrophysics and Cosmology, Department of Physics, and  \\
        SLAC National Accelerator Laboratory, Stanford University, Stanford, CA 94305, USA \\
        }

\date{Accepted 2013 xx xx;  Received 2013 xx xx; in original form 2013 xx xx}

\pagerange{\pageref{firstpage}--\pageref{lastpage}} \pubyear{2013}

\def\LaTeX{L\kern-.36em\raise.3ex\hbox{a}\kern-.15em
    T\kern-.1667em\lower.7ex\hbox{E}\kern-.125emX}

\begin{document}

\label{firstpage}

\maketitle

\begin{abstract}
\emph{XMM-Newton} observations of the supergiant fast X-ray transient IGR~J17544$-$2619 are reported and placed in the context of an analysis of archival \emph{INTEGRAL}/IBIS data that provides a refined estimate of the orbital period at 4.9272$\pm$0.0004\,days. A complete outburst history across the \emph{INTEGRAL} mission is reported. Although the new \emph{XMM-Newton} observations (each lasting $\sim$15\,ks) targeted the peak flux in the phase-folded hard X-ray light curve of IGR~J17544$-$2619, no bright outbursts were observed, the source spending the majority of the exposure at intermediate luminosities of the order of several 10$^{33}$\,erg\,s$^{-1}$ (0.5\,$-$\,10\,keV) and displaying only low level flickering activity. For the final portion of the exposure, the luminosity of IGR~J17544$-$2619 dropped to $\sim$4$\times$10$^{32}$\,erg\,s$^{-1}$ (0.5 - 10 keV), comparable with the lowest luminosities ever detected from this source, despite the observations being taken near to periastron. We consider the possible orbital geometry of IGR~J17544$-$2619 and the implications for the nature of the mass transfer and accretion mechanisms for both IGR~J17544$-$2619 and the SFXT population. We conclude that accretion under the `quasi-spherical accretion' model provides a good description of the behaviour of IGR~J17544$-$2619, and suggest an additional mechanism for generating outbursts based upon the mass accumulation rate in the hot shell (atmosphere) that forms around the NS under the quasi-spherical formulation. Hence we hope to aid in explaining the varied outburst behaviours observed across the SFXT population with a consistent underlying physical model.

\end{abstract}


\section{Introduction}

Supergiant Fast X-ray Transients (SFXT) are a sub-class of supergiant high mass X-ray binaries (HMXB) that display extreme flaring behaviour on short ($\sim$ hour) timescales \citep{2005A&A...444..221S}. They also display an X-ray dynamic range in excess of that possessed by classical supergiant X-ray binaries ($\sim$\,10$-$20, Sg-XRB), reaching 10$^{4}$\,$-$\,10$^5$ in the most extreme systems such as XTE J1739$-$302 \citep{2006ApJ...638..974S} whilst only reaching about 10$^2$ in the so-called intermediate SFXTs, such as IGR J16465$-$4514 \citep{2010MNRAS.406L..75C}. Due to the supergiant nature of their companion stars \citep{2006ESASP.604..165N} SFXTs are located along the Galactic Plane where there are currently 10 spectroscopically confirmed systems \citep{2011arXiv1111.5747S}, as well as a similar number of candidate SFXTs that show the required X-ray flaring behaviour but, as yet, do not have a spectroscopically confirmed supergiant counterpart. 

IGR J17544$-$2619 was first discovered as a hard X-ray transient source on 2003 September 17 \citep{2003ATel..190....1S} with the IBIS/ISGRI instrument aboard \emph{INTEGRAL}. Two short outbursts, 2 and 8 hours long respectively, were observed on the same day, indicating fast and recurrent transient behaviour. After the \emph{INTEGRAL} detection,  IGR J17544$-$2619 was associated with the soft X-ray source 1RXS J175428.3-262035 (\citealt{2003ATel..191....1W}, \citealt{2000IAUC.7432....3V}). A \emph{Chandra} observation \citep{2005A&A...441L...1I} precisely located the source with a positional accuracy of 0.6'' (RA = 17:54:25.284, DEC = -26:19:52.62, J2000.0), confirming the association of IGR J17544$-$2619 with the IR source 2MASS J17542527-2619526. \citet{2006A&A...455..653P} classified the companion as an O9Ib star with a mass of 25$-$28\,M$_{\odot}$ at a distance of 2$-$4\,kpc.  Subsequently \citet{2008A&A...484..801R} performed SED fitting to the mid-IR spectrum and refined the distance estimate for the system to $\sim$3.6\,kpc. 

Soft X-ray observations of IGR J17544$-$2619 have illustrated the large scale variability and different emission states that the source can display and inhabit. In its lowest quiescent state, IGR J17544$-$2619 displays a soft, likely thermal spectrum (e.g. $\Gamma$\,$=$\,$5.9\pm1.2$, \citealt{2005A&A...441L...1I}) and luminosities of down to $\sim$8$\times$10$^{31}$\,erg\,s$^{-1}$ \citep{2004A&A...420..589G}. Conversely, during bright outbursts the source displays hard powerlaw spectra with luminosities of up to several 10$^{36}$\,erg\,s$^{-1}$ and significant variability in the spectral parameters, particularly absorption, over the duration of the event (e.g.  \citealt{2009ApJ...707..243R}). Both the deep quiescent and bright outburst states are, however, relatively rare in IGR J17544$-$2619, comprising only a few percent of the system duty cycle each, while a spectrally hard intermediate emission state, with X-ray luminosity in the range 10$^{33}$\,$-$\,10$^{34}$\,erg\,s$^{-1}$, is most commonly inhabited by IGR J17544$-$2619 \citep{2011MNRAS.410.1825R}. As a result of this X-ray behaviour, the recurrent hard X-ray outbursts and its spectrally classified 09Ib companion, IGR J17544$-$2619 is often referred to as the `prototypical' SFXT.

Using long baseline IBIS/ISGRI light curves, \citet{2009MNRAS.399L.113C} identified the orbital period of IGR J17544$-$2619 as 4.926 $\pm$ 0.001\,d, one of the shortest orbital periods observed in an SFXT. This orbital period was utilised to place constraints on the binary orbit with an upper limit of 0.4 placed on the eccentricity if the companion possesses its minimum possible radius of 12.7\,$R_{\odot}$ (\citealt{2006A&A...455..653P}, \citealt{2008A&A...484..801R}). Additionally \citet{2012A&A...539A..21D} detected X-ray pulsations with a period of 71.49$\pm$0.02\,s from the region of IGR J17544$-$2619 using the non-imaging \emph{RXTE}/PCA instrument. By considering the known source population within the field of view, they concluded that IGR J17544$-$2619 was the most likely origin of the pulsed emission. The compact object in IGR J17544$-$2619 is almost certainly a neutron star and such pulsed emission would confirm that. When combined with the orbital period of \citet{2009MNRAS.399L.113C}, the pulse period places the system at a location consistent with the classical SgXRBs in the $P_{\rm orb}$\,$-$\,$P_{\rm spin}$ parameter space of the Corbet diagram \citep{1986MNRAS.220.1047C}.

In this work we present new orbital phase targeted \emph{XMM-Newton} observations of IGR J17544$-$2619, along with an archival \emph{INTEGRAL} study of the system. In Section \ref{sect:INTanal} we outline the data set, analysis and results from the archival \emph{INTEGRAL} observations. Section \ref{sect:XMManal} then presents the analysis of, and results from, the new \emph{XMM-Newton} data sets. Finally these results are discussed in Section \ref{sect:discuss} and final conclusions drawn in Section \ref{sect:conc}.

\label{sect:intro}

\section{Archival \emph{INTEGRAL} data analysis and results}

The archival \emph{INTEGRAL}/IBIS (\citealt{2003A&A...411L...1W}, \citealt{2003A&A...411L.131U}) data set consisted of all observations of IGR J17544$-$2619 spanning 2003 November 25 through 2012 Feburary 10, providing a total exposure of $\sim$16.9\,Ms. All observations were processed with version 9 of the \emph{INTEGRAL} Offline Science Analysis software (OSA, \citealt{2003A&A...411L.223G}) and images were created in the 18$-$60\,keV energy band for each science window (ScW, individual \emph{INTEGRAL} exposures of nominal length 2000\,s). An 18$-$60\,keV light curve was generated by extracting count rates and errors from each image at the best determined X-ray position of IGR J17544$-$2619. 

\begin{figure}
\begin{center}
\includegraphics[width=0.5\textwidth]{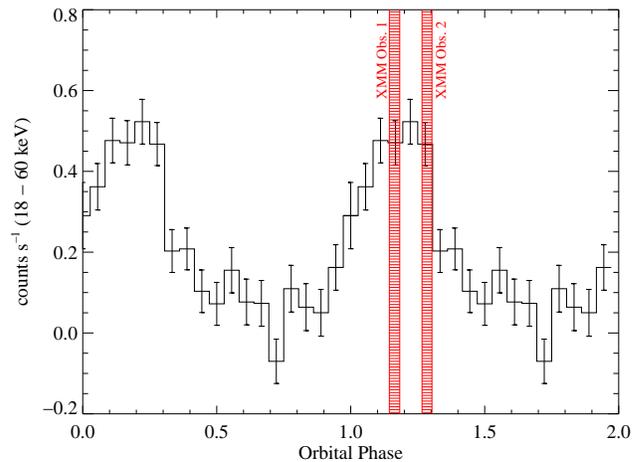}
\caption{\emph{INTEGRAL}/IBIS 18$-$60\,keV orbital phase-folded light curve of IGR J17544$-$2619, using the orbital period of 4.9272\,days and a zero phase ephemeris of MJD 55924.271. The shaded vertical bands illustrate the regions of orbital phase covered by the new \emph{XMM-Newton} observations.}
\label{fig:pfoldedlc}
\end{center}
\end{figure}

The light curve was filtered for observations where the total exposure time was less than 200\,s and where IGR J17544$-$2619 was located at an off-axis angle of greater than 12$^{o}$, to remove data points with the largest intrinsic uncertainty. The filtered light curve was then tested for the presence of periodicities by the production of a Lomb-Scargle periodogram (\citealt{1976Ap&SS..39..447L}, \citealt{1982ApJ...263..835S}). Using the statistical tests outlined in \citet{2013MNRAS.433..528D}, the orbital period of IGR J17544$-$2619 was significantly detected at 4.9272$\pm$0.0004\,days. This orbital period is consistent with, and provides a refined determination of, the value reported by \citet{2009MNRAS.399L.113C} and will be used throughout the remainder of this work. Figure \ref{fig:pfoldedlc} shows the orbital phase-folded light curve of IGR J17544$-$2619, produced using the new period determination and a zero-phase ephemeris of MJD 55924.271. IGR J17544-2619 displays a smooth, but asymmetric orbital flux profile, with the active region covering a phase interval of $\Delta\phi\sim$\,0.4, corresponding to a duration $\sim$2.0\,days. Outside of this active region, there is little significant emission detected from IGR J17544$-$2619. Additionally the orbital phase coverage of the new \emph{XMM-Newton} observations are shown by the shaded vertical regions (see Section \ref{sect:XMManal} below for further details). We note that the \emph{XMM-Newton} non-detection reported by \cite{2004A&A...420..589G} occurred at phase $0.85 \pm 0.05$, while the two detections in the same work, and the peak flare in \cite{2009ApJ...707..243R} occurred at phases of $0.23 \pm 0.05$, $0.16 \pm 0.05$ and $0.97 \pm 0.01$ respectively. These are all consistent with the folded light curve presented in Figure~\ref{fig:pfoldedlc}.

The archival 18 $-$ 60\,keV light curve was also searched for outbursts with durations in the range 0.02 to 2.0\,days ($\sim$half an hour to just under half of one orbit) using the method outlined in \citet{2013MNRAS.433..528D}. This analysis identified 31 distinct outburst events within the archival IBIS light curve, down to a minimal significance of 4.5$\sigma$ and having durations in the range $\sim$0.3 to 20.9\,hours. While 14 of the 31 detected outbursts have been reported in previous works, the full set of outbursts and their properties are reported in Table \ref{tab:IBIS_bursts} to provide  a complete history of the outburst activity of IGR J17544$-$2619 as seen by \emph{INTEGRAL}/IBIS. The orbital phases reported in Table \ref{tab:IBIS_bursts} relate to the ScW with the maximal count rate in each outburst event, calculated using the orbital period determination and ephemeris outlined above. The flux of each outburst in Table \ref{tab:IBIS_bursts} is calculated using PIMMS to convert the peak ISGRI count rate detected during each outburst, assuming a thermal Bremsstrahlung model (BREMMS) with a plasma temperature of 10\,keV. This model is defined by the best fit to the spectrum of the most significant outburst of IGR J17544$-$2619 detected by IBIS ($\hat{\chi}^{2} =$\,0.78\,(4 $d.o.f.$), achieved using XPSEC v12.7.1 \citep{1996ASPC..101...17A}). Unfortunately, however, there were insufficient statistics to perform a full spectral study of all the outbursts detected by IBIS.

\begin{figure}
\begin{center}
\includegraphics[width=0.5\textwidth]{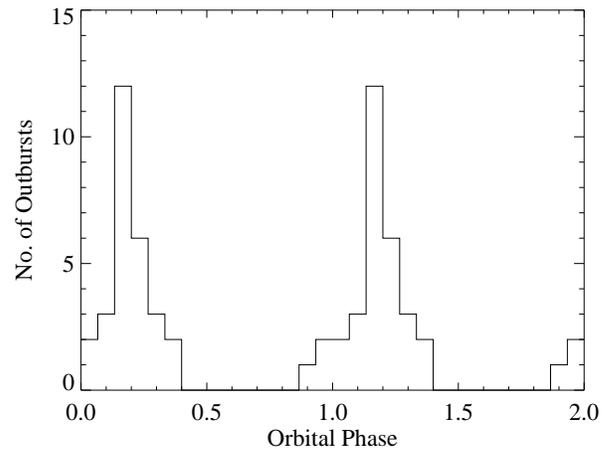}
\caption{Orbital phase distribution of the complete set of IGR J17544$-$2619 outbursts detected by \emph{INTEGRAL}/IBIS in the 18$-$60\,keV band using the same orbital ephemeris as Fig. \ref{fig:pfoldedlc}. The distribution is seen to peak in the same region of orbital phase as the phase-folded light curve.}
\label{fig:pfoldedbursts}
\end{center}
\end{figure}

The total combined duration of the detected outbursts of IGR J17544$-$2619 is $\sim$225\,ks, representing an outburst duty cycle of $\sim$1.3\% in the 18$-$60\,keV band. This is broadly consistent with the value of 0.5\% derived by \citet{2010MNRAS.408.1540D} who performed a systematic study in the 20$-$40\,keV band on a data set of similar total exposure (for a discussion of the origin of the discrepancy between the duty cycle derived from each analysis see \citet{2013MNRAS.433..528D}). Using the orbital period and zero-phase ephemeris outlined above, the orbital phase of the peak of each outburst event was calculated and the orbital phase distribution of outbursts constructed. This distribution is shown in Fig. \ref{fig:pfoldedbursts} and it is seen that the outburst locations are peaked across the same region of orbital phase as the phase-folded light curve. Both of these plots suggest that the periodicity detected in IGR J17544$-$2619 is being driven by the eccentric motion of the neutron star around the supergiant companion that preferentially induces X-ray emission during the periastron passage of the neutron star. This is in contrast to the SFXTs with shorter orbital periods, whereby the observed periodicity is driven by the eclipse of the NS in a more circular orbit (i.e. IGR J16418$-$4514, \citealt{2006ATel..779....1C}). Further discussions on the nature of the physical processes driving the X-ray emission observed from IGR J17544$-$2619 are given in Section \ref{sect:discuss}.

A new \emph{INTEGRAL}/IBIS data set was also obtained simultaneously to the \emph{XMM-Newton} observations described in Section \ref{sect:XMManal}. The continuous observations were performed between 2012-09-16 01:16:58 and 2012-09-17 01:09:06, covering a phase range of $\phi =$\,0.130 to 0.332, for a total exposure of 83\,ks. IGR J17544$-$2619 was not detected during this observation with a 3$\sigma$ upper limit of 0.5 counts per second, corresponding to a flux of 2.0$\times$10$^{-11}$\,erg\,cm$^{-2}$\,s$^{-1}$ (18$-$60\,keV), assuming the same spectral model as used for the outburst flux conversions in Table \ref{tab:IBIS_bursts}. Despite targeting the peak of the phase-folded light curve, the non-detection in this particular observation is consistent with the duty cycle derived from the archival observations. 

\begin{table*}
	\begin{center}
	\caption{Complete outburst history of IGR J17544$-$2619 identified in the archival \emph{INTEGRAL}/IBIS dataset. The orbital phase given relates to the ScW with the maximal count rate in each event using the orbital period determination $P_{\rm orb}$\,$=$\,4.9272\,days and $t_{\rm zero}$\,$=$\,MJD 55924.271. The peak flux is that of the same ScW in the 18\,$-$\,60\,keV band, calculated assuming a thermal Bremsstrahlung model with a temperature of 10\,keV as measured from the spectrum of the most significant outburst detected.}
	\begin{tabular}{|c|c|c|c|c|c|}
	\hline
	\multicolumn{1}{|c|}{Significance} & \multicolumn{1}{|c|}{Start MJD} &  \multicolumn{1}{|c|}{End MJD} & \multicolumn{1}{|c|}{Duration} & \multicolumn{1}{|c|}{Orbital Phase} & \multicolumn{1}{|c|}{Peak Flux} \\
	 & & & (hours) & $\phi$ & erg\,cm$^{-2}$\,s$^{-1}$ \\ \hline
	6.0\phantom{-} & 52732.35 & 52732.39 & 0.9 & 0.18 & 2.8$\times$10$^{-10}$ \\
	29.7$^{a}$ & 52899.05 & 52899.58 & 12.6 & $\sim$0.00 & 9.7$\times$10$^{-10}$ \\
	5.0\phantom{-} & 52904.01 & 52904.58 & 13.6 & $\sim$0.15 & 2.3$\times$10$^{-10}$ \\
	5.3$^{b}$ & 53062.34 & 53063.21 & 20.9 & $\sim$0.2 & 4.6$\times$10$^{-10}$ \\
	6.0\phantom{-} & 53067.28 & 53067.49 & 5.1 & $\sim$0.2 & 2.3$\times$10$^{-10}$ \\
	26.0$^{c}$ & 53072.36 & 53072.50 & 3.5 & 0.21 & 1.4$\times$10$^{-9}$ \\
	11.6$^{c}$ & 53072.54 & 53072.70 & 4.1 & 0.23 & 4.6$\times$10$^{-10}$ \\
	9.9$^{d}$ & 53269.89 & 53270.08 & 4.5 & 0.31 & 3.8$\times$10$^{-10}$ \\
	5.1\phantom{-} & 53422.34 & 53422.36 & 0.6 & 0.22 & 5.1$\times$10$^{-10}$ \\
	8.7$^{d}$ & 53441.23 & 53441.25 & 0.4 & 0.06 & 5.1$\times$10$^{-10}$ \\
	5.1\phantom{-} & 53451.41 & 53451.50 & 2.0 & 0.14 & 2.5$\times$10$^{-10}$ \\
	10.7$^{b}$ & 53481.26 & 53481.33 & 1.6 & 0.18 & 3.9$\times$10$^{-10}$ \\
	5.9$^{b}$ & 53628.73 & 53628.87 & 3.3 & 0.12 & 3.6$\times$10$^{-10}$ \\
	11.7$^{*}$\phantom{-} & 53658.35 & 53658.36 & 0.3 & 0.12 & 6.5$\times$10$^{-10}$ \\
	5.8$^{*}$\phantom{-} & 53658.37 & 53658.41 & 0.9 & 0.13 & 2.4$\times$10$^{-10}$ \\
	16.0$^{b}$ & 53806.87 & 53807.04 & 4.1 & 0.27 & 6.8$\times$10$^{-10}$ \\
	5.6\phantom{-} & 53850.87 & 53851.11 & 2.7 & 0.24 & 4.5$\times$10$^{-10}$ \\
	8.0$^{b}$ & 53987.52 & 53987.59 & 1.6 & 0.94 & 3.2$\times$10$^{-10}$ \\
	11.3$^{b}$ & 53998.42 & 53998.60 & 4.3 & 0.14 & 9.9$\times$10$^{-10}$ \\
	9.6\phantom{-} & 54343.56 & 54343.63 & 1.6 & 0.20 & 4.1$\times$10$^{-10}$ \\
	11.4$^{e}$ & 54364.25 & 54364.32 & 1.9 & 0.39 & 5.0$\times$10$^{-10}$ \\
	11.1\phantom{-} & 54367.23 & 54367.25 & 0.6 & 0.99 & 5.2$\times$10$^{-10}$ \\
	4.8\phantom{-} & 54545.89 & 54545.95 & 1.5 & 0.26 & 2.0$\times$10$^{-10}$ \\
	20.3$^{*}$\phantom{-} & 54560.24 & 54560.36 & 2.7 & 0.17 & 1.1$\times$10$^{-9}$ \\
	4.5$^{*}$\phantom{-} & 54560.37 & 54560.63 & 6.1 & $\sim$0.2 & 1.9$\times$10$^{-10}$ \\
	5.4$^{e}$ & 54570.44 & 54570.45 & 0.2 & 0.23 & 5.6$\times$10$^{-10}$ \\
	8.8$^{e}$ & 54708.09 & 54708.17 & 1.9 & 0.18 & 4.3$\times$10$^{-10}$ \\
	5.2\phantom{-} & 54886.35 & 54886.42 & 1.6 & 0.36 & 1.9$\times$10$^{-10}$ \\
	5.8\phantom{-} & 54938.38 & 54938.39 & 0.3 & 0.91 & 3.7$\times$10$^{-10}$ \\
	6.8\phantom{-} & 55067.64 & 55067.96 & 7.6 & $\sim$0.15 & 2.5$\times$10$^{-10}$ \\
	12.4\phantom{-} & 55437.96 & 55438.02 & 1.4 & 0.30 & 5.7$\times$10$^{-10}$ \\ 
	\hline
	\end{tabular}
	\label{tab:IBIS_bursts}
	\end{center}
	\begin{flushleft}
	{\footnotesize Note: marked outbursts have previously been reported in the following works: $^{a}$\citealt{2003ATel..190....1S}; $^{b}$\citealt{2009MNRAS.399L.113C}; $^{c}$\citealt{2004ATel..252....1G}; $^{d}$\citealt{2006ApJ...646..452S}; $^{e}$\citealt{2007A&A...466..595K}.$^{*}$Likely individual flares within the same outburst event.}
	\end{flushleft}
\end{table*}

\label{sect:INTanal}

\section{\emph{XMM-Newton} data analysis and results}

Two separate \emph{XMM-Newton}/EPIC (\citealt{2001A&A...365L...1J}, \citealt{2001A&A...365L..27T}, \citealt{2001A&A...365L..18S}) observations of IGR J17544$-$2619 were performed that targeted the periastron region of the systems orbital phase. The first observation was performed between UTC 02:55:14 and 07:30:14 2012-09-16, while the second observation was performed between UTC 17:20:14 and 21:46:54 2012-09-16, with each exposure totalling $\sim$15\,ks. The first and second observations covered the orbital phase range $\phi$\,$=$\,0.144\,$-$\,0.183 and 0.266\,$-$\,0.303 respectively, as shown by the vertical shaded areas of Fig. \ref{fig:pfoldedlc}. As can be seen in Fig. \ref{fig:pfoldedlc} both observations occurred during the peak of the phase-folded light curve, however, no significant hard X-ray emission was detected from IGR J17544$-$2619 in the simultaneous \emph{INTEGRAL}/IBIS observations.

\begin{figure}
\begin{center}
\includegraphics[width=0.5\textwidth]{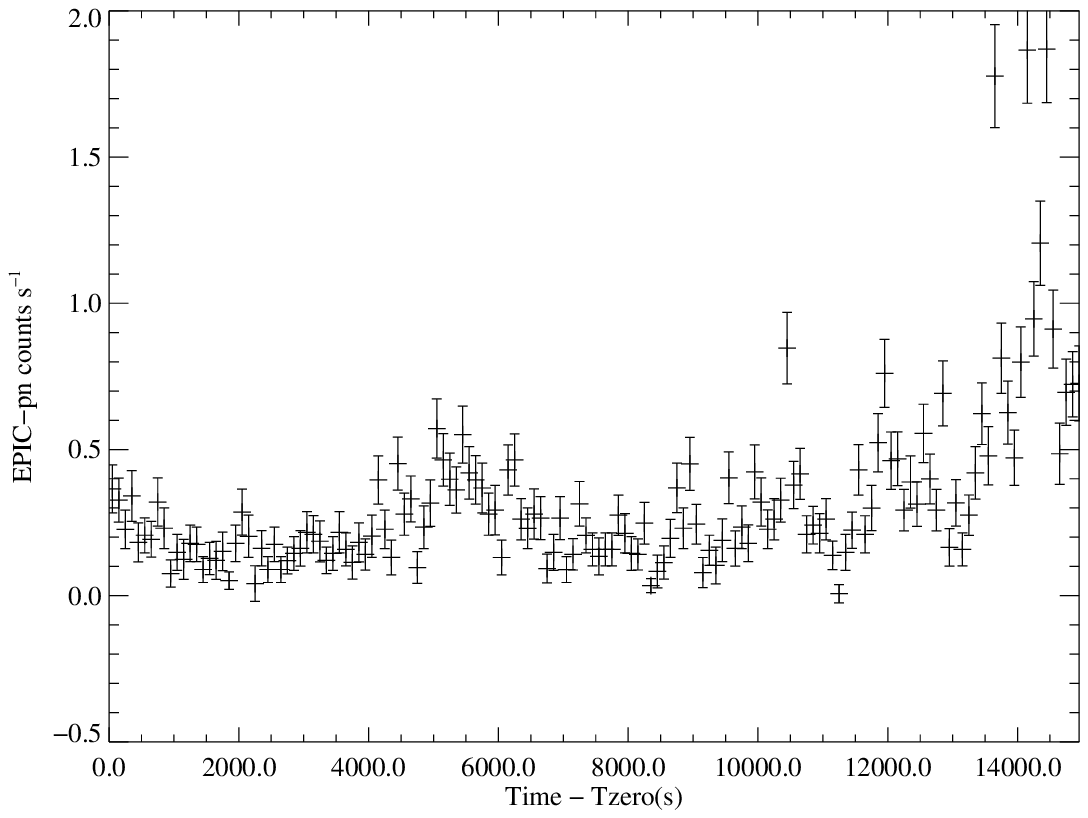}
\includegraphics[width=0.5\textwidth]{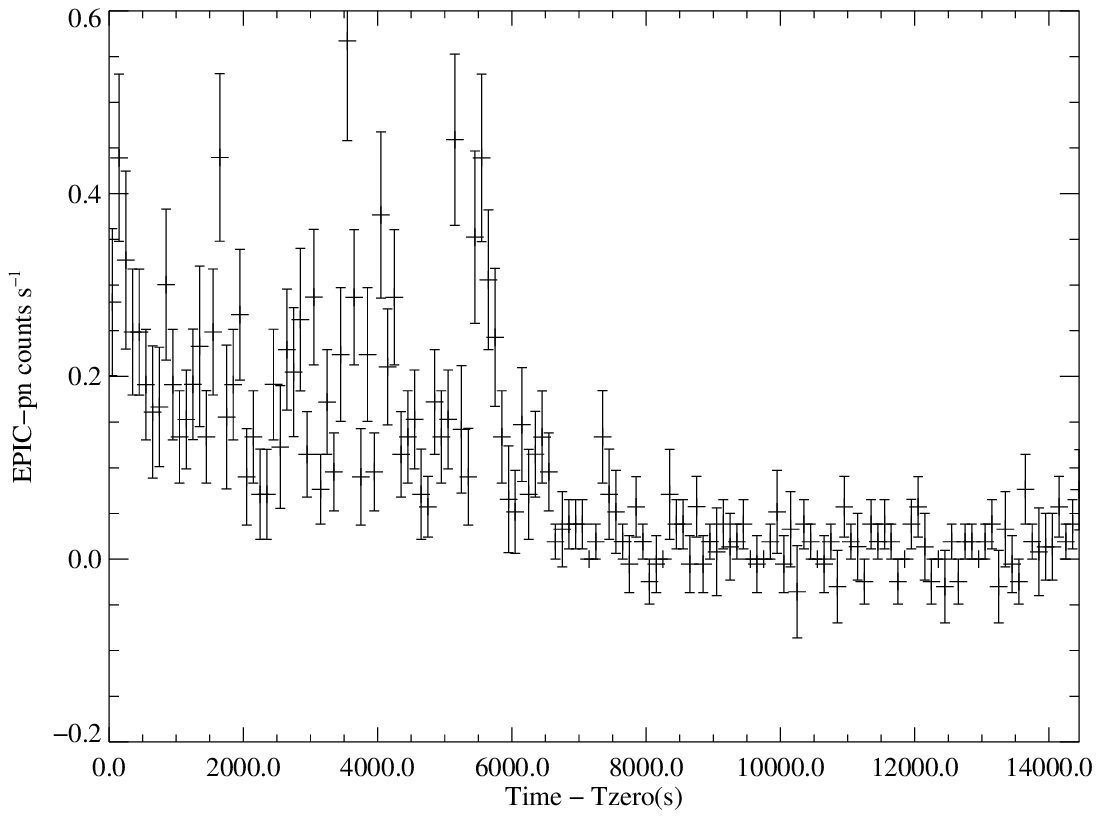}
\caption{0.2$-$10\,keV EPIC-pn light curves of the first (top) and second (bottom) \emph{XMM-Newton} observations of IGR J17544$-$2619 at 100\,s resolution. In each case $t_{\rm zero}$ corresponds to the first time stamp in the pn light curve.}
\label{fig:XMM_lcs}
\end{center}
\end{figure}

Data from both the EPIC-MOS and EPIC-pn detectors were analysed for each observation using SAS v12.0.1 \citep{2004ASPC..314..759G} and the most recent instrument calibration files. The data sets were filtered for regions of high particle background and photon pile-up following the methods outlined in the \emph{XMM-Newton} SAS data analysis threads\footnote{http://xmm.esac.esa.int/sas/current/documentation/threads/}. Each observation was found to be unaffected by both particle flaring and photon pile-up, allowing the full exposure length to be used for scientific data product extraction. Images were extracted from both the MOS and pn cameras, with IGR J17544$-$2619 being detected at a similar low flux level in each. Due to the low signal-to-noise obtained in the MOS cameras, however, it was not possible to extract further meaningful data products. Hence only the EPIC-pn data is considered further in this work. The EREGIONANALYSE tool was used to define the optimal extraction region for all light curve and spectral generation procedures performed during this study. Using extraction regions of radius 25 and 17 arcseconds for the first and second observations respectively, the broadband 0.2$-$10\,keV EPIC-pn light curves of each observation were extracted at 100 second resolution, as shown in Fig. \ref{fig:XMM_lcs}. The properties of these light curves are discussed in Section \ref{sect:XMMtemp} and the spectral analysis performed on the data set outlined in Section \ref{sect:XMMspec}. All spectra reported in this work were extracted following the standard procedures and the SAS tools RMFGEN and ARFGEN were used to extracted the necessary response files for each spectrum. The spectra were again fit using XSPEC version 12.7.1, with uncertainties quoted at the 90\% confidence level throughout and the elemental abundances set to those of \citet{2000ApJ...542..914W}. 

\label{sect:XMManal}

\subsection{Temporal Analysis}

Figure \ref{fig:XMM_lcs} shows the variations in the soft X-ray flux detected during the first (top) and second (bottom) \emph{XMM-Newton} observations. In both cases $t_{\rm zero}$ relates to the first time stamp of the EPIC-pn light curve of that observation. In the first observation the source is observed to be active at a low base level, with a small increase in flux between $t=$\,4500$-$6500\,s, followed by increased flickering activity at times t\,$>$\,10,000\,s. While the first observation shows relatively smooth variations in flux within an active state, the second observation shows two distinct regions of emission. At times $t<$\,7000\,s IGR J17544$-$2619 shows flickering activity consistent with the flux level of the first observation, both in temporal variability and flux range. At later times, however, the source drops into a low state with very little detectable emission and no evidence of flickering. This evolution is intriguing and may signify the switching of accretion modes during this observation, further consideration of which is given in Section \ref{sect:discuss}.

Due to the low signal-to-noise of the light curves, however, little additional temporal information could be extracted from the data. Hardness ratio light curves drawn from the 0.2\,$-$\,4 / 4\,$-$\,10\,keV bands did not show any significant variability due to the intrinsic uncertainty on the data points, especially in the regions of lowest flux. Additionally the weak strength of the detections did not allow the production of finely binned (e.g. 1\,s) light curves to allow thorough testing for the proposed 71.49\,s pulse period of IGR J17544$-$2619 \citep{2012A&A...539A..21D}. Light curves from both observations were extracted with larger 10\,s bins and tested for periodicities via the Lomb-Scargle method, however, the intrinsic uncertainties of the light curve data points resulted in noise dominated periodograms devoid of any significant signals and contaminated by red noise due to the longer time scale variations observed in both exposures. The data were also tested for periodicities at the event level by means of an epoch folding analysis, again, however, no significant signals were detected at any frequency in either observation. The lack of detectable, coherent pulsations within these light curves is intriguing and suggests, despite the poor statistics, that the proposed 71.49\,s pulsation detected with \emph{RXTE}/PCA should be treated with caution. Further discussion of the properties of the \emph{RXTE} detection and \emph{XMM-Newton} non-detection data sets, along with the implications of the 71.49\,s signal not belonging to IGR J17544$-$2619, are given in Section \ref{sect:discuss}.

\label{sect:XMMtemp}

\subsection{Spectral Analysis}

The EPIC-pn spectra from the full exposure of each observation were extracted using optimally determined EREGIONANALYSE circular regions, with radii of 25 and 17 arcseconds for the first and second observations respectively, and binned to a minimum of 15 counts per bin. The spectra were fit in the 0.5\,$-$\,15\,keV energy band within XSPEC, and were both observed to be smooth with no indications of emission lines. In both cases, however, the total spectrum was poorly described by simple absorbed models such as those using black body ({\it phabs(bbody)}) or powerlaw ({\it phabs(powerlaw)}) continua. Instead the use of Comptonisation models and/or the extraction of spectra from different regions of the exposure was required to achieve statistically satisfactory fits to the data. Due to the differing spectral fitting procedures performed for each observation, the specific analysis performed for each are discussed in turn in the following paragraphs, while a summary of the fitted models and parameters is shown in Table \ref{tab:XMM_specparam}.   

\begin{table*}
\begin{center}
\caption{Best fitted spectral models of the exposure regions outlined in Section \ref{sect:XMMspec}. The first two lines show the fits to the active exposure regions using a Comptonised black body model ({\it phabs(compbb)}) while the next two lines show the fits to the same regions using a simple absorbed powerlaw ({\it phabs(powerlaw)}). Finally the last line shows the fit to the low flux region using a simple absorbed powerlaw, with the $n_{\rm H}$ fixed to the Galactic value, achieved using C-statistics$^{*}$ due to the low signal-to-noise of the extracted spectrum. The spectral parameter shown for the fits to the total exposure of the first observation relate to the fit and residuals shown in Fig. \ref{fig:XMMObs1_multiresid} while the parameters of the low flux region relate to Fig. \ref{fig:XMMObs2_quiescentspec}.}
\begin{tabular}{|c|c|c|c|c|c|c|}
\hline
\multicolumn{1}{|c|}{Section} & \multicolumn{1}{|c|}{$n_{\rm H}$} & \multicolumn{1}{|c|}{kT$_{\rm bb}$} & \multicolumn{1}{|c|}{$\tau$} & \multicolumn{1}{|c|}{$\Gamma$} & \multicolumn{1}{|c|}{Absorbed flux (0.5\,$-$\,10\,keV)} & \multicolumn{1}{|c|}{$\hat{\chi^{2}}$ ({\it dof})} \\
 & 10$^{22}$\,cm$^{-2}$ & keV & & & $\times$10$^{-12}$\,erg\,cm$^{-2}$\,s$^{-1}$ & \\ \hline
Obs 1 - Total & 1.4$^{+0.25}_{-0.21}$ & 0.79$^{+0.08}_{-0.08}$ & 1.3$^{+0.20}_{-0.20}$ & $-$ & 2.06$^{+0.05}_{-0.55}$ & 1.04 (101) \\ 
Obs 2 - Active & 1.2$^{+0.50}_{-0.38}$ & 0.86$^{+0.15}_{-0.14}$ & 0.7$^{+0.38}_{-0.48}$ & $-$ & 1.13$^{+0.06}_{-0.48}$ & 0.77 (41) \\ \hline
Obs 1 - Total & 2.7$^{+0.32}_{-0.28}$ & $-$ & $-$ & 1.9$^{+0.12}_{-0.12}$ & 2.08$^{+0.08}_{-0.09}$ & 1.28 (102) \\
Obs 2 - Active & 3.1$^{+0.73}_{-0.61}$ & $-$ & $-$ & 2.3$^{+0.28}_{-0.26}$ & 1.15$^{+0.08}_{-0.20}$ & 0.99 (42) \\
Obs 2 - Low flux & 1.44 (fixed) & $-$ & $-$ & 2.1$^{+0.40}_{-0.37}$ & 0.12$^{+0.024}_{-0.028}$ & 3.16 (4)$^{*}$ \\ \hline
\end{tabular}
\label{tab:XMM_specparam}
\end{center}
\end{table*}

\begin{figure}
\begin{center}
\includegraphics[width=0.5\textwidth]{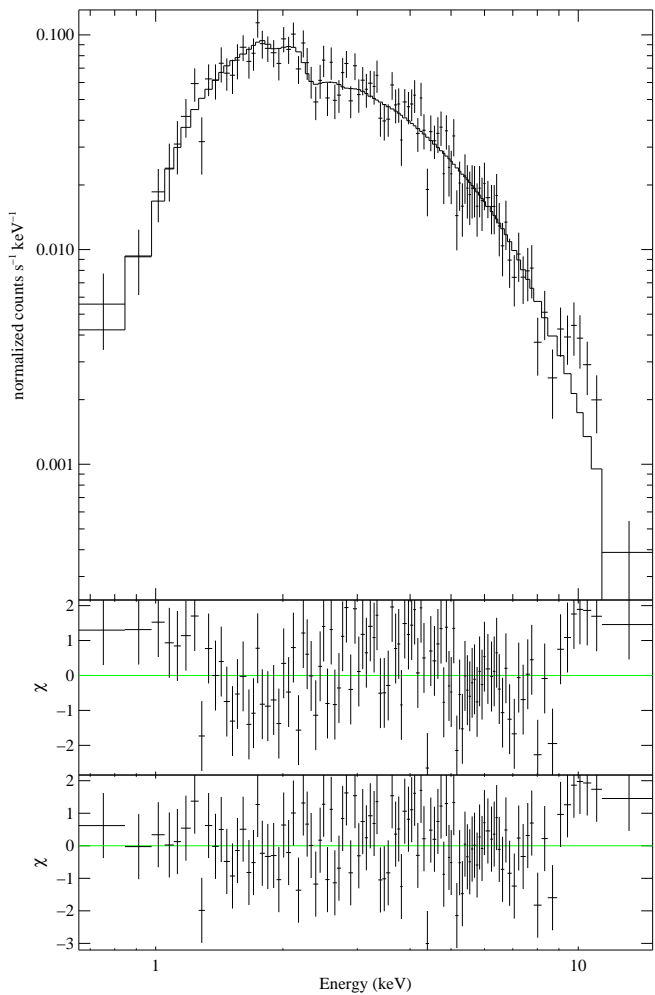}
\caption{EPIC-pn spectrum extracted from the full exposure of the first observation. The top panel shows the spectrum and model while the bottom panel shows the residuals for the best fit Comptonised black body model, whose parameters are shown in Table \ref{tab:XMM_specparam}. For comparison the residuals of the fit to the spectrum using a simple absorbed powerlaw model are shown in the middle panel.}
\label{fig:XMMObs1_multiresid}
\end{center}
\end{figure}

As noted above, the spectrum extracted from the full exposure of the first observation could not be described by a simple absorbed powerlaw, with the best fit achieved $\hat{\chi}^{2}$ having a value of 1.28 (102 {\it dof}). The absorption was initially set to the Galactic value in the direction of IGR J17544$-$2619 ($n_{\rm H}$\,$=$\,1.44$\times$10$^{22}$\, cm$^{-2}$, \citealt{1990ARA&A..28..215D}), and then left as a free parameter, and the residuals of the resulting best fit are shown in the middle panel of Fig. \ref{fig:XMMObs1_multiresid}. Instead the spectrum could be well described by a more physical Comptonised black body model, whereby the seed photons produced in the accretion columns are subsequently Compton scattered by the hot plasma contained there ({\it phabs(compbb)}, \citealt{1986PASJ...38..819N}). As shown in the top row of Table \ref{tab:XMM_specparam} and the residuals in the lower panel of Fig. \ref{fig:XMMObs1_multiresid}, the spectrum was well fit by this model and displayed an absorption consistent with the Galactic value ($n_{\rm H}$\,$=$\,(1.4$^{+0.25}_{-0.21}$)$\times$10$^{22}$\,cm$^{-2}$), a blackbody temperature of kT$_{\rm bb}$\,$=$\,0.79$^{+0.08}_{-0.08}$\,keV and an optical depth of $\tau$\,$=$\,1.3$^{+0.20}_{-0.20}$ (note, following the standard convention for this model, the electron temperature of the Comptonising plasma kT$_{\rm e}$ is fixed at 50\,keV and the optical depth allowed to vary to avoid degeneracies in these parameters). The average flux detected during the first observation was (2.06$^{+0.05}_{-0.55}$)$\times$10$^{-12}$\,erg\,cm$^{-2}$\,s$^{-1}$ (0.5$-$10\,keV). As there is no scientific requirement for IGR J17544$-$2619 to have remained in a single spectral state throughout the observation, however, spectra were also extracted from smaller regions of the exposure and fit with the same Comptonisation model to investigate any variations in the spectral parameters throughout the observation. The four regions used corresponded to the start of the observation (t $<$ 4500 in the top panel of Fig. \ref{fig:XMM_lcs}), the flux `bump' (4500 $\leq$ t $<$ 6500), the intermediate section (6500 $\leq$ t $<$ 13,500) and the final flaring (t $\geq$ 13,500) section of the light curve. Despite the fact that well defined spectra were extracted from these regions, no significant variations in the spectral parameters were observed across the observation, with the absorption, black body temperature and optical depth being consistent, within uncertainties, with the values obtained from the total exposure. 

\begin{figure}
\begin{center}
\includegraphics[angle=270,width=0.5\textwidth]{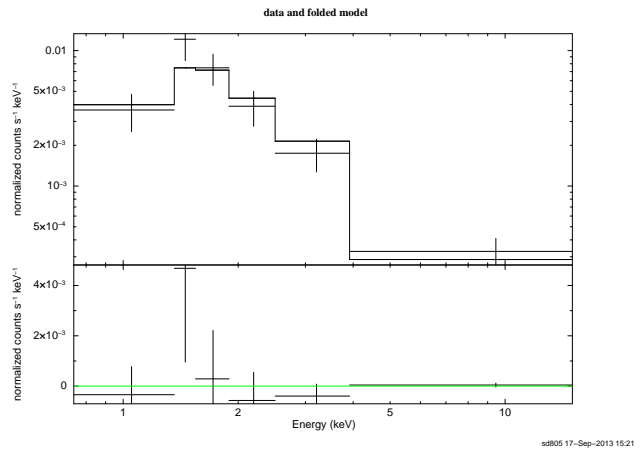}
\caption{EPIC-pn spectrum extracted from the low flux region of the second observation and fit with a simple absorbed powerlaw model using C-statistics to account for the low signal-to-noise.}
\label{fig:XMMObs2_quiescentspec}
\end{center}
\end{figure}

In contrast to the first observation, however, the total spectrum extracted from the second observation could not be well fit by either the absorbed powerlaw or Comptonised black body models. Instead, given the varied behaviour observed in the light curve of the second observation (see Fig. \ref{fig:XMM_lcs}), two separate spectra were extracted for the active (t $<$ 7000\,s) and low flux (t $\geq$ 7000\,s) portions of the light curve. In this case, however, the spectrum of the active region could be well fit by both the powerlaw ($\hat{\chi^{2}}$\,$=$\,0.99 (42 {\it dof})) and the Comptonised black body ($\hat{\chi^{2}}$\,$=$\,0.77 (41 {\it dof})) models. In both cases the derived spectral parameters were broadly consistent with the values obtained from the first observation, as shown in Table \ref{tab:XMM_specparam}, with the powerlaw model displaying an absorption slightly in excess of the Galactic value in the direction of IGR J17544$-$2619 and the Comptonised model displaying an absorption consistent with the Galactic value. The average flux derived from the powerlaw fit of the active region of the second observation was (1.15$^{+0.08}_{-0.20}$)$\times$10$^{-12}$\,erg\,cm$^{-2}$\,s$^{-1}$ (0.5 - 10\,keV). The spectrum extracted from the low flux region of the light curve did not, however, contain enough signal-to-noise to perform spectral fitting using the standard Chi-squared statistic, being comprised of only $\sim$100 photons, and instead required the use of the `C-statistic' and the fixing of the absorption to the Galactic value. Furthermore the spectrum was only fit with a simple powerlaw model and displayed a photon index of 2.1$^{+0.4}_{-0.4}$ and a flux of 1.2$\times$10$^{-13}$\,erg\,cm$^{-2}$\,s$^{-1}$ (0.5\,$-$\,10\,keV), which also corresponds to the data, model and residuals shown in Fig. \ref{fig:XMMObs2_quiescentspec}. Due to the very limited statistics accumulated during this region of the exposure, however, more physical models (such as neutron star atmospheres (NSA)) could not be applied to the spectrum of the low flux region. 

Through comparing the parameter sets shown in Table \ref{tab:XMM_specparam} it is seen that it is difficult to draw conclusions on the level of spectral variability observed across these exposures, especially given that a single model does not provide a satisfactory description of all three extracted spectra. We note, however, that the flux detected from the source in the low flux state is comparable to the lowest previously reported detection of IGR J17544$-$2619 \citep{2005A&A...441L...1I} which, given the associated change in the temporal behaviour of the light curve, may indicate that IGR J17544$-$2619 has entered a true quiescent state at this time, despite the fact that the spectrum is not as soft as the previous quiescence detection ($\Gamma$\,=\,2.1 in this work and $5.9 \pm 1.2$ (90\% confidence) in the \emph{Chandra} observation reported by \citet{2005A&A...441L...1I}). In this case the observed X-ray emission may potentially originate from the hot NS surface, the NS atmosphere or the intrinsic X-ray emission of the companion star itself (e.g. \citealt{1997A&A...322..167B}). While, as stated above, the statistics are insufficient to perform a full spectral investigation of these origins, a fuller discussion of the nature of the low flux state is given in Section \ref{sect:discuss_nature}.   
       
\label{sect:XMMspec}

\section{Discussion}

The observations described in this work have utilised the combined capabilities of \emph{INTEGRAL} and \emph{XMM-Newton} to further the characterisation of the prototypical SFXT IGR J17544$-$2619. Using the data products extracted from these observations we now consider the orbital geometry of the system and the accretion mechanisms active within it, leading us to propose a possible new outburst generation mechanism in SFXTs. 

For consistency we use our newly enhanced orbital period, derived from the archival \emph{INTEGRAL}/IBIS data, to place constraints on the geometry of IGR J17544$-$2619s orbit. Assuming the system contains a NS with a nominal mass of 1.4\,$M_{\rm \odot}$ and a supergiant with a mass at the lower end of the range proposed by \citet{2006A&A...455..653P}, 25\,$M_{\rm \odot}$, the semi-major axis of the 4.9272\,day binary orbit is calculated as 36.3\,$R_{\rm \odot}$. Using the $R_{\rm \star}$\,$/$\,$D_{\rm \star}$ SED parameter value for IGR J17544$-$2619 and the refined distance estimate of 3.6\,kpc \citep{2008A&A...484..801R} also implies a stellar radius for the supergiant of 20.3\,$R_{\rm \odot}$. Under these assumptions eccentricities of up to $\sim$0.3 are dynamically valid using the constraints that Roche-Lobe overflow does not occur in IGR J17544$-$2619 and that the supergiant, L1 point separation is given by the relationship of \citet{1964BAICz..15..165P}. The constraint that RLO does not occur in IGR J17544$-$2619 is supported by the difference in the behaviours observed between this system and systems powered by RLO, which display bright ($\sim$10$^{38}$\,erg\,s$^{-1}$) persistent X-ray emission and signatures of the accretion discs present, such as super orbital modulation generated by the precession of a warped accretion disc (e.g. SMC X-1, \citealt{1998ApJ...502..253W}) or the X-ray dips generated by the flared edge of an accretion disc viewed at an approximately edge on inclination (e.g. Cen X-3, \citealt{2011ApJ...737...79N}). 

The range of eccentricities allowed by the above assumptions are required to explain the properties of the orbital phase folded IBIS light curve and outburst distribution displayed in Section \ref{sect:INTanal}. The peaked shape of the phase folded profiles shown in both Fig. \ref{fig:pfoldedlc} and Fig. \ref{fig:pfoldedbursts} illustrate that it is the preferential generation of enhanced emission at certain orbital phases that generates the periodic signal in the archival IGR J17544$-$2619 data. The preferentially generated emission is most likely related to the periastron passage of a NS which possesses a significant orbital eccentricity, as opposed to the effects of varying line of sight absorptions as the NS orbits its companion, which are insignificant in the hard X-ray band utilised by IBIS. This eccentricity also represents a different behaviour to that observed from the two SFXTs with shorter orbital periods, IGR J16479$-$4514 \citep{2009MNRAS.397L..11J} and IGR J16418$-$4514 \citep{2006ATel..779....1C}, where the periodic modulation is generated as a result of X-ray eclipses within almost circular orbits (e.g. \citealt{2013MNRAS.433..528D}). It should be noted, however, that while in the nomenclature used here the peak in X-ray flux is referred to as periastron, it is not possible to evaluate how this equates to the physical closest approach of the NS without achieving a dynamically derived orbital solution for IGR J17544$-$2619. While the phase-folded outburst distribution shown in Fig. \ref{fig:pfoldedbursts} illustrates that the outburst events are preferentially generated near the periastron passage of the NS, it is not yet clear what drives the underlying modulation of the hard X-ray flux in IGR J17544$-$2619. This modulation is still present if the bright outburst events are removed from the light curve (see \citealt{2009MNRAS.399L.113C}) and may be generated by, either, a modulation of the flux across the orbital phase as a result of a smoothly varying stellar wind environment, or the superposition of multiple smaller flares that illustrate an underlying, phase dependent, flare production probability distribution. However, given the \emph{Swift}/XRT monitoring results of \citet{2011MNRAS.410.1825R} and the orbital phase location of the low flux region detected in the new \emph{XMM-Newton} observations (Fig. \ref{fig:pfoldedlc}), it appears that a superposition of distinct flaring events may be the more likely emission mechanism that generates the orbital flux modulation observed in IGR J17544$-$2619 by IBIS.

While the \emph{INTEGRAL} data can be explained by the presence of a significant eccentricity in the orbit of IGR J17544$-$2619, the \emph{XMM-Newton} observations provide a challenge to the invocation of orbital eccentricity as the driving factor behind production of the deep quiescent states observed in SFXTs (e.g. \citealt{2008AIPC.1010..235C}). While the wider orbits of the SFXTs with the longest orbital periods (e.g. the $\sim$30\,day orbit of SAX J1818.6$-$1703, \citealt{2009MNRAS.393L..11B}) may facilitate the inhabiting of a quiescent state at apastron separations (although continuity considerations suggest the full dynamic range of SFXTs cannot be totally explained this way), the detection of a deep quiescence in a system with a short orbital period and with only a small separation from the presumed periastron passage, illustrates that additional processes must be responsible for the generation of such events. We now, therefore, consider the higher order accretion mechanisms that may be active in IGR J17544$-$2619 in the context of explaining the various emission states observed by \emph{XMM-Newton}.    

\label{sect:discuss}

\subsection{The nature of the active accretion mechanism of IGR J17544$-$2619}

To consider the active accretion regimes that may be operating in IGR J17544$-$2619 we calculate the X-ray luminosity of the three sections of the \emph{XMM-Newton} light curve assuming a distance of 3.6\,kpc \citep{2008A&A...484..801R}. The average unabsorbed 0.5$-$10\,keV luminosities of the whole first observation, the active region of the second observation and the low flux region of the second observation were 6.7$\times$10$^{33}$, 5.3$\times$10$^{33}$ and 3.7$\times$10$^{32}$\,erg\,s$^{-1}$ respectively (calculated using WebPIMMS and the powerlaw spectral parameters and fluxes reported in Table \ref{tab:XMM_specparam}). Additionally the outburst population described in Table \ref{tab:IBIS_bursts} displays 18$-$60\,keV X-ray luminosities in the range 2.9$\times$10$^{35}$ to 2.2$\times$10$^{36}$\,erg\,s$^{-1}$. This set of luminosities broadly covers the full range of emission states observed in SFXTs, namely the bright outbursts ($\sim$10$^{36}$\,erg\,s$^{-1}$), the spectrally hard intermediate state (10$^{33}$\,$-$\,10$^{34}$\,erg\,s$^{-1}$) and the spectrally soft quiescent state ($\sim$10$^{32}$\,erg\,s$^{-1}$). 

It is, however, difficult to conclude as to whether IGR J17544$-$2619 entered a true quiescent state, whereby accretion on to the surface of the NS has ceased, at the end of the second \emph{XMM-Newton} observation as, whilst the unabsorbed luminosity is comparable to that of the quiescence reported by \citet{2005A&A...441L...1I}, the photon index of the powerlaw does not increase to the thermal soft state reported in that work (e.g. $\Gamma =$\,$5.9\pm 1.2$). Instead, in this case, the photon index of the powerlaw fit to the spectrum of the quiescent state remains consistent with that observed from the earlier active regions, suggesting that accretion may have continued at a very low level during this time. However, the temporal behaviour of the emission during this time is markedly different as shown in the lower panel of Fig. \ref{fig:XMM_lcs}, suggesting that a switching off of accretion may have occurred. In this scenario the residual emission results from either, the hot NS surface, the NS atmosphere or the stellar wind of the companion star. Whilst there are some spectral models that describe some of these processes (e.g. the NSA model, \citealt{1996A&A...315..141Z}), there were insufficient statistics to perform detailed fitting of the spectrum of the low flux region to provide a direct diagnostic that IGR J17544$-$2619 entered a true quiescence. Given the rarity, and the intrinsic low quality of all spectra extracted from  quiescent episodes of SFXTs (the spectrum extracted from the \emph{Chandra} observation reported by \citet{2005A&A...441L...1I} comprised only 26 photons), it is difficult to conclude on the exact nature, and level of variability, of quiescence in SFXTs. One possible origin of the discrepancy in the spectral shape and the slightly higher luminosity observed during these observations compared to the previous quiescence detection (3.7$\times$10$^{32}$ as opposed to 1.8$\times$10$^{32}$\,erg\,s$^{-1}$), however, is that the low flux state was directly preceded by a period of active accretion. Hence even if the accretion was cut-off at the end of the active region, the accretion columns/mounds may still have been cooling throughout the final region of the observation, leading to the higher flux and harder spectrum when compared to the \citet{2005A&A...441L...1I} quiescence, which was detected at the start of their observations such that the system may have inhabited that state for an extended period prior to the observation. 

Insights into the wider nature of the system can be generated by evaluating the conditions under which the action of the `Propeller effect' \citep{1975A&A....39..185I}, which inhibits the accretion of material due to the centrifugal force generated by the rotating NS, becomes dominant. The propeller effect becomes active when the rotation of the NS magnetosphere becomes supersonic with respect to the in-falling stellar wind material and results in captured material being ejected from the NS, removing angular momentum. Conceptually this process occurs at the point where the radius of the NS magnetosphere ($R_{\rm M}$, also known as the Alven radius) becomes equal to the co-rotation radius ($R_{\rm co}$, the radius where the rotational velocity of the magnetosphere is equal to the local Keplerian velocity). While $R_{\rm co}$ is simply defined through circular motion and is consistent across all models, $R_{\rm M}$ varies depending upon the specific accretion model invoked, the specific luminosity of the central source, the external stellar wind conditions and the intrinsic magnetic field strength of the NS. 

For the considerations presented here we utilise the recent `Quasi-spherical accretion' model of \citet{2012MNRAS.420..216S}, which invokes a hot shell of gravitationally captured stellar wind plasma (analogous to an atmosphere) about the NS to mediate the rate of accretion on the NS surface through different dominant plasma cooling mechanisms at the base of the shell. Under this model gravitationally captured stellar wind material cannot enter the NS magnetosphere directly, and instead requires plasma at the base of the shell to be cooled and enter the magnetosphere through plasma instabilities, which occur most efficiently in the equatorial regions of the NS magnetosphere. As discussed further in \citet{2013MNRAS.428..670S}, at X-ray luminosities of below $\sim$3$\times$10$^{35}$\,erg\,s$^{-1}$ the NS accretion columns are optically thin, resulting in the formation of pencil beams that only illuminate the magnetosphere polar regions. In this configuration plasma at the base of the hot shell can only cool via radiative processes (e.g. Bremsstrahlung) resulting in low accretion rates. If a system is at an X-ray luminosity of greater than $\sim$3$\times$10$^{35}$\,erg\,s$^{-1}$, however, the accretion columns are optically thick, generating fan beam emission that illuminates the equatorial region and cools the plasma through more efficient Compton processes, generating higher X-ray luminosities. We chose this model in particular due to its ability to explain both the SFXTs and the persistent SgXRBs under a consistent theoretical framework. As outlined in \citet{2013MNRAS.428..670S}, it may be the case that the difference between the two populations is the proportion of time spent in each cooling regime, with the persistent sources naturally inhabiting the higher luminosity Compton-cooled regime whilst the SFXTs naturally spend the majority of the time in the lower luminosity radiatively-cooled regime, but occasionally transition into the Compton-cooled regime during the bright outbursts. 

This interpretation of \citet{2013MNRAS.428..670S} is supported by the luminosity of the different types of emission observed from IGR J17544$-$2619 described at the beginning of this section and in previous observing campaigns, which agree well with the transition luminosities of the quasi-spherical accretion model (e.g. bright outbursts detected by \emph{INTEGRAL}/IBIS have luminosities of $\geq$\,3$\times$10$^{35}$\,erg\,s$^{-1}$ while the most commonly inhabited intermediate luminosity state displays luminosities in the region of several 10$^{33}$ to 10$^{34}$\,erg\,s$^{-1}$ \citep{2011MNRAS.410.1825R}). 

The same is confirmed also by the source luminosities reported in the literature by \citet{2004A&A...420..589G} and by \citet{2009ApJ...707..243R}. Calculating the source luminosities at a distance of 3.6 kpc (instead of 8 kpc, as incorrectly assumed by \cite{2004A&A...420..589G}), we find that during the low level flaring activity observed by XMM-Newton in September 2003 the 0.5-10 keV luminosity ranged between 1.4$\times$10$^{33}$~erg~s$^{-1}$ and 1.7$\times$10$^{35}$~erg~s$^{-1}$, confirming a low level accretion dominated by inefficient radiative processes, as predicted by Shakura et al. (2012). Converting the 2-10 keV fluxes reported by \cite{2009ApJ...707..243R} using WebPIMMS (adopting their quoted power law parameters), into 0.5-10 keV luminosities, we found that the source, during this long Suzaku observation, always displayed a luminosity lower than 10$^{35}$~erg~s$^{-1}$, except during the brightest flare (where it reached 4$\times$10$^{36}$~erg~s$^{-1}$), again consistent with the \cite{2012MNRAS.420..216S} model, where a much higher luminosity is allowed by the Compton-cooled regime above 3$\times$10$^{35}$~erg~s$^{-1}$.

Additionally, the NS shell can act as a damping mechanism that negates the effects of the variability of the supergiant stellar winds which should, according to the hydrodynamic simulations of \citet{2012MNRAS.421.2820O}, generate variations in X-ray luminosity of up to eight orders of magnitude if the emission produced directly traced the variations in the local stellar wind conditions encountered by the NS. We therefore apply the formulations of this model to investigate the nature of the low flux state observed below, before discussing a possible implication of the quasi-spherical accretion model on the mechanism that produces SFXT outbursts in the following section.

Under the quasi-spherical accretion model, the radius of the NS magnetosphere can be defined using Equation 17 of \citet{2013MNRAS.428..670S}, namely:

\begin{equation}
	R_{M} = 10^{9} L_{36}^{-6/27} \mu_{30}^{16/27} \textnormal{cm}
\end{equation}

{\noindent where $L_{36}$ is the X-ray luminosity in units of 10$^{36}$\,erg\,s$^{-1}$ and $\mu_{30}$ is the NS magnetic moment ($\mu = V_{NS} B_{NS}$) in units of 10$^{30}$\,G\,cm$^{3}$. The co-rotation radius can be defined as:}

\begin{equation}
	R_{co} = 1.7\times10^{10} P_{s3}^{2/3} \textnormal{cm}
\end{equation}

{\noindent where $P_{s3}$ is the spin period of the NS in units of 1000\,s \citep{2008ApJ...683.1031B}. Equating the above expressions allows the NS magnetic field strength required to facilitate the entry into the propellor regime at a given luminosity to be calculated as:}

\begin{equation}
	\mu_{30} = ( 17 P_{s3}^{2/3} L_{36}^{6/27} )^{27/16}
	\label{eq:prop}
\end{equation}

{\noindent Using the derived luminosity of the quiescent state (3.7$\times$10$^{32}$\,erg\,s$^{-1}$) with the suggested NS spin period of IGR J17544$-$2619 (71.49\,s, \citealt{2012A&A...539A..21D}), however, yields a magnetic field strength of only $\sim$8$\times$10$^{10}$\,G which is significantly below the nominal B-field strengths of HMXB pulsars. In this case, this would suggest that the quiescent states observed in IGR J17544$-$2619 are not generated through the sudden cessation of accretion, but are instead very low luminosity accreting states. Such a scenario is somewhat supported by the lack of significant variability of the powerlaw photon index between the two regions of the second \emph{XMM-Newton} observation, and the fact that the NSA model fitted to the quiescent spectrum observed by \emph{Chandra} was only done so using the zero magnetic field variant of the model \citep{2005A&A...441L...1I}, albeit the strength of the conclusions that can be drawn from both studies are severely limited by poor statistics. Alternatively, we might suspect that the 71.49\,s signal detected by \citet{2012A&A...539A..21D} did not originate from IGR J17544$-$2619 (see \citealt{2012A&A...539A..21D} for a full discussion of the source population within the \emph{RXTE}/PCA FOV at the time of this detection) and that the system instead contains a more slowly rotating NS as observed in some other SFXTs (e.g. 1246s for IGR J16418$-$4532, \citealt{2006A&A...453..133W}). In this case,  the B-field required to balance Eq. \ref{eq:prop} for the observed luminosity could exceed $\sim 10^{12}$\,G. This value is more consistent with the magnetic field strengths measured in HMXBs that display cyclotron lines (e.g. 4U 1907$+$09, \citealt{2013arXiv1309.0875H}) and, taken with the fact that no periodic modulations were detected in the \emph{XMM-Newton} observations, may suggest that the origin of the 71.49\,s periodicity was misidentified by \citet{2012A&A...539A..21D}.}

With the \emph{XMM-Newton} observations presented in this work it is difficult to make a definitive judgement as to the true nature of the quiescent states observed in SFXTs. However, combining the lack of a detection of the previously reported pulse period of IGR J17544$-$2619 in these observations, with the fact that a cessation in accretion can be generated for a typical NS magnetic field in IGR J17544$-$2619 if the NS possessed a longer spin period, suggests that the 71.49\,s pulse period of IGR J17544$-$2619 should be treated with caution until further observations with focusing instruments have either confirmed or rejected the existence of such a pulsation. In particular, observations of IGR J17544$-$2619 using focusing instruments with the source at a comparable luminosity to the observation in which the 71.49\,s pulsation was detected ($\sim$10$^{34}$\,erg\,s$^{-1}$) would be highly informative in assessing the origin of this signal. 

In either case, however, we believe that the quasi-spherical accretion model \citep{2012MNRAS.420..216S} provides one of the best explanations for generating the observed behaviour in both the SFXT and persistent SgXRB populations. By considering the implications on the state of the NS shell in a system which can accrete in regimes of different efficiency, we propose a new mechanism for triggering outbursts in SFXTs under this model as outlined in the following section.  

\label{sect:discuss_nature}

\subsection{A new method of SFXT outburst generation $-$ the `accumulation mechanism'}

In their further discussions of the `quasi-spherical accretion' model \citet{2013MNRAS.428..670S} have suggested that the difference between SFXTs and the classical persistent SgXRBs, such as Vela X-1, may be that SFXTs preferentially inhabit the lower luminosity, radiatively cooled accretion regime for the majority of the time, and only enter the higher luminosity Compton cooled regime for brief periods. These brief periods are detected as the rapid X-ray outbursts that characterise the SFXT class. To generate the transitions \citet{2013MNRAS.428..670S} invoke the action of density increases above the magnetosphere, analogous to the clumpy-wind model (\citealt{2005A&A...441L...1I}, \citealt{2007A&A...476..335W}), to temporarily increase the accretion rate, associated X-ray luminosity and activate the Compton cooling mechanism. We note, however, that a brief increase in the density above the magnetosphere may be a specific example of a more general consequence of prolonged periods of accretion in the in-efficient radiatively cooled regime. 

As discussed above the Compton and radiative cooled accretion modes within the quasi-spherical accretion model have different efficiencies which, in both cases, can be interpreted as $\eta = \dot{M}_{\rm acc} / \dot{M}_{\rm capt}$, where $\dot{M}_{\rm acc}$ and $\dot{M}_{\rm capt}$ are the time dependent accretion and mass capture rates respectively. The implication of this relationship is that, while $\eta <$\,1, matter is accumulating within the NS shell as it is captured from the stellar wind at a higher rate than it is accreted on to the NS surface via plasma instabilities at the base of the NS shell. For a system inhabiting the radiatively-cooled regime, the rate of radiative cooling is simply dependent upon the density of plasma at the base of the shell, such that prolonged periods of matter accumulation will increase the accretion rate achieved through radiative cooling. As the matter continues to accumulate the X-ray luminosity generated through the accretion of the radiatively cooled plasma will therefore also continue to increase, until the point that the critical X-ray luminosity is achieved and the Compton cooling mechanism is activated. At this point the accretion rate, and X-ray luminosity, will increase rapidly as a result of the increased efficiency of the plasma cooling at the base of the magnetosphere. Under this interpretation, therefore, the driving mechanism in producing an outburst in an SFXT can simply be thought of as the point at which a sufficient amount of matter has accumulated in the NS shell such that the X-ray luminosity produced by radiative cooling at the base of the shell is sufficient to generate a switch in to the Compton cooled regime. 

The advantage of this mechanism is that SFXT-like outbursts could be generated in systems with a wide range of orbital and stellar wind parameters. For example, outbursts could, in theory, be generated in systems that possessed circular orbits and a smooth stellar wind as, over time, the density of the shell would increase, eventually triggering a regime switch. In reality, however, the stellar winds of supergiant stars are known to be structured (e.g. \citealt{1988ApJ...335..914O}, \citealt{2008AJ....136..548L}) such that it is inevitable that $\dot{M}_{\rm capt}$ will vary as a function of time. Under this mechanism, however, such variations can simply be parameterised as an associated time variability of $\eta$ and the mass accumulation rate in the shell. Similarly, the effects of the significant eccentricities that have been inferred in the orbits of many SFXTs, including IGR J17544$-$2619 in this work, can also be parameterised by a variation in $\eta$ as a function of orbital phase. Therefore such an `accumulation mechanism' has the potential to explain the individual behaviours observed from both the short orbital period, near circular systems (e.g. IGR J16479$-$4514, \citealt{2009MNRAS.397L..11J}) and the longer orbital period, likely eccentric systems (e.g. SAX J1818.6$-$1703, \citealt{2009MNRAS.393L..11B}) through only the difference in the time dependence of $\eta$ in each case. 

It should be noted also, that this accumulation mechanism is not mutually exclusive from the clumpy-wind theory that has been long associated with the SFXT class. If a NS encounters a particularly dense clump then the rate of accumulation in the shell may be sufficiently rapid as to generate a switch to the Compton-cooled regime in a short period of time, such that the clump directly drives the outburst. In addition to this, however, the accumulation mechanism could also generate an outburst after a NS has spent a long period of time accreting from a relatively smooth, lower density stellar wind (e.g. the apastron passages of long orbital period systems such as SAX J1818.6$-$1703). The accumulation mechanism also naturally acts as a damping mechanism to prevent the observationally unsupported 8 orders of magnitude variations predicted in all SgXRBs by the hydrodynamic simulations of \citet{2012MNRAS.421.2820O} if the X-ray luminosity was to directly trace the local mass capture rate in such systems. Additionally the accumulation mechanism removes the need for all encounters with over dense regions of the stellar wind to generate bright outbursts, which in turn may help to define the low outburst duty cycles, $\sim$1\,$-$\,2\%, observed in most SFXTs \citep{2010MNRAS.408.1540D}.   

Finally, the accumulation mechanism may also provide insights into the length and temporal properties of SFXT outbursts. As the systems can only inhabit the higher efficiency Compton-cooled state while there is sufficient accretion to power an X-ray source with a luminosity of $\sim$3$\times$10$^{35}$\,erg\,s$^{-1}$, they will only be able to maintain this higher efficiency accretion whilst there is sufficient matter at the base of the NS shell. The duration of the outbursts may therefore be defined by the time required to `drain' the shell of a sufficient amount of the accumulated material such that the Compton cooling can no longer produce a high enough accretion rate to maintain the higher efficiency accretion, and the source returns to the lower luminosity radiatively cooled regime. Furthermore, if the characteristic accumulation timescale is longer than the orbital period of the binary, this mechanism may be an important factor in producing the, currently un-explained, super orbital periodicities recently detected in the X-ray light curves of some SgXRBs, including the SFXTs IGR J16418$-$4532 and IGR J16479$-$4514, by both \emph{Swift}/BAT and \emph{INTEGRAL}/IBIS (\citealt{2013arXiv1309.4119C}, \citealt{2013ATel.5131....1D} respectively).  

The `accumulation mechanism' described in this section seeks to utilise the recently reported quasi-spherical accretion model of \citet{2012MNRAS.420..216S} to describe a base mechanism through which the outbursting behaviour of SFXTs can be generated across the wide range of orbital parameters reported for these systems ($P_{\rm orb}$ in the range $\sim$3 to 165\,days (\citealt{2009MNRAS.397L..11J}, \citealt{2007A&A...476.1307S})), while also removing the requirement of the X-ray luminosity directly tracing the mass capture rate (i.e. stellar wind velocity and density). Whilst this mechanism may be able to aide in explaining the observed SFXT phenomenology, further in-depth numerical modelling is required to allow for precise predictions and conclusions to be drawn.         

\section{Conclusions}

This work has presented a new multi-epoch \emph{XMM-Newton} study of the SFXT IGR J17544$-$2619, along with an archival \emph{INTEGRAL}/IBIS study of the source. IGR J17544$-$2619 was observed to be active at a low level, with small flaring events, across the majority of the \emph{XMM-Newton} exposure. In the final region of the \emph{XMM-Newton} exposure, however, IGR J17544$-$2619 was detected in one of the lowest flux states observed from this source. This state displayed markedly different temporal behaviour from the earlier active regions of the EPIC light curves. Significant variations in the spectral parameters between the different flux states could not be observed, however, although the lack of signal-to-noise in the low flux region prevented a detailed spectral classification of this state. By considering the luminosity evolution observed in the new \emph{XMM-Newton} observations in the context of the previous activity of the source, and the orbital phase parameters provided by the \emph{INTEGRAL} data, we discuss the nature of the observed emission states within the context of the recently proposed `quasi-spherical accretion' model \citep{2012MNRAS.420..216S}. We conclude that IGR J17544$-$2619 is likely inhabiting the lower luminosity, radiatively-cooled regime of the quasi-spherical accretion model during the new \emph{XMM-Newton} observations. We also caution that, under this model, the previously reported 71.49\,s pulse period may have been misidentified as originating from IGR J17544$-$2619 if the low flux state represents a true quiescence whereby accretion is completely impeded by a centrifugal barrier. We note, however, that further detailed studies of the quiescent states of SFXTs, along with additional searches for the proposed pulse period, are required to make a definitive judgement on the nature of the quiescent states observed in IGR J17544$-$2619. 

Following these conclusions, we outline a possible new mechanism of SFXT outburst generation that utilises the rate of accumulation of plasma in a NS atmosphere to trigger a switch into the more efficient Compton-cooled accretion regime of the quasi-spherical accretion model. This new mechanism seeks to unify the varying behaviours of different SFXT systems through the interpretation of the variations as individual time dependences of the accretion efficiency $\eta = \dot{M}_{\rm acc} / \dot{M}_{\rm capt}$ in each case. In addition to attempting to provide a unified underlying physical cause of producing SFXT outbursts across the diverse range of orbital geometry possessed by these systems, this `accumulation mechanism' also removes the requirement of SFXT luminosity variations to be a direct tracer of the local properties of the supergiant stellar wind, a scenario which has posed serious problems to the interpretation of SFXT observations. To further the characterisation of this possible new outburst generation mechanism, we encourage detailed numerical modelling of the accumulation scenario, whilst also encouraging continued multi-waveband follow-up of the SFXTs. In particular the achievement of dynamically derived orbital solutions of SFXT orbits (through either radial velocity studies in the optical/IR or through pulsar timing) and the characterisation of the supergiant stellar wind properties (through the modelling of emission line profiles in UV where possible) would greatly aide in the characterisation of the accretion regimes inhabited by SFXTs.         

\label{sect:conc}

\section*{Acknowledgements}

The authors would like to thank I. Negueruela for his helpful comments on the paper. This work was based on observations with \emph{INTEGRAL}, an ESA project with instruments and science data centre funded by ESA member states (especially the PI countries: Denmark, France, Germany, Italy, Switzerland, Spain) and with the participation of Russia and the USA. This work was also based on observations obtained with \emph{XMM-Newton}, an ESA science mission with instruments and contributions directly funded by ESA Member States and NASA. S. P. Drave acknowledges support from the UK Science and Technology Facilities Council, STFC. M. E. Goossens is supported by a Mayflower scholarship from the University of Southampton. A. Bazzano acknowledges financial support from ASI/INAF agreement n. 2013-025.R.0. A. B. Hill acknowledges that this research was supported by a Marie Curie International Outgoing Fellowship within the 7$^{th}$ European Community Framework Programme (FP7/2007--2013) under grant no. 275861. This research has made use of the SIMBAD database, operated at CDS, Strasbourg, France. This research has made use of the IGR Sources page maintained by J. Rodriguez \& A. Bodaghee (http://irfu.cea.fr/Sap/IGR-Sources/).  


\label{lastpage}

\end{document}